\documentclass[preprint,12pt]{aastex}               
\usepackage{epsfig}

\shortauthors{D.~A.~Uzdensky}

\begin{document}

\title{Force-Free Magnetosphere of an Accretion Disk --- Black Hole
System. I. Schwarzschild Geometry
\thanks{KITP preprint NSF-KITP-03-85}}
\author{Dmitri A. Uzdensky}
\affil{Kavli Institute for Theoretical Physics, University of California} 
\affil{Santa Barbara, CA 93106}
\email{uzdensky@kitp.ucsb.edu}
\date{October 12, 2003}

\begin{abstract}

In this paper I study the magnetosphere of a black hole that is 
connected by the magnetic field to a thin conducting Keplerian 
disk. I consider the case of a Schwarzschild black hole only, 
leaving the more interesting but difficult case of a Kerr black 
hole to a future study. I assume that the magnetosphere is ideal, 
stationary, axisymmetric, and force-free. I pay a special attention 
to the two singular surfaces present in the system, i.e., the event 
horizon and the inner light cylinder; I use the regularity condition 
at the light cylinder to determine the poloidal electric current
as a function of poloidal magnetic flux. I solve numerically the 
Grad--Shafranov equation, which governs the structure of the 
magnetosphere, for two cases: the case of a nonrotating disk
and the case of a Keplerian disk. I find that, in both cases,
the poloidal flux function on the horizon matches a simple 
analytical expression corresponding to a radial magnetic field
that is uniform on the horizon. Using this result, I express 
the poloidal current as an explicit function of the flux and 
find a perfect agreement between this analytical expression 
and my numerical results.

\end{abstract}



\section{Introduction}
\label{sec-intro}

It has been broadly acknowledged that magnetic fields around
accreting black holes are very important. Magnetic interaction 
between a spinning black hole and remote astrophysical loads 
is often invoked to explain many observed features of Active 
Galactic Nuclei (AGNs) and Galactic black holes (e.g., Begelman, 
Blandford, \& Rees 1984; Krolik~1999; Punsly~2001). In particular, 
magnetic configurations in which the field lines threading a black 
hole extend to infinity have been studied very extensively and have 
gained a lot of popularity as the standard model for jet production. 
In this model, the black hole's rotational energy is extracted 
electromagnetically by the means of the famous Blandford--Znajek 
mechanism (Blandford \& Znajek 1977, hereafter BZ77; Macdonald 
\& Thorne 1982, hereafter MT82; Phinney~1983; Macdonald~1984; 
Thorne~et~al. 1986; Komissarov~2001) and is transported outward
in the form of Poynting flux to power a jet.

Recently, however, another magnetic configuration has become
a subject of growing interest --- a configuration where at 
least some part of magnetic field lines connect the black hole 
and the accretion disk (e.g., MT82; Nitta~et~al. 1991; Hirotani
et~al. 1992; Blandford~1999, 2000; Gruzinov~1999; Li~2000, 
2001, 2002; Wang~et~al. 2002, 2003). In this so-called 
Magnetically-Coupled (MC) configuration (Wang~et~al. 2002)
the magnetic field couples the hole directly to the disk 
and transfers angular momentum between the two; it can thus 
regulate the spin evolution of the black hole (Wang~et~al. 2002, 
2003). In addition, the magnetic link provides a means to extract 
the rotational energy of the black hole (in a manner similar 
to the~BZ77 mechanism) and to transport it to the inner region 
of the disk. This effect may lead to some additional heating 
and an increase in the luminosity of the inner part of the disk, 
with important observational implications (Gammie~1999; Li 2000, 
2001, 2002). Another reason for interest in the~MC configuration 
is the suggestion that twisted (due to a mismatch of rotation 
rates of the hole and the disk) field lines may become unstable, 
leading to strong variability on the rotation time scale and 
possibly to Quasi-Periodic Oscillations (QPOs), as suggested 
by Gruzinov (1999). QPOs may also be produced by non-axisymmetries 
of the magnetic field connecting the hole to the disk and the 
associated non-axisymmetry of local disk heating (Li~2001, 2002).

In theoretical studies of both the BZ77 and MC processes, researchers 
(including Blandford \& Znajek themselves) have often employed the 
framework of force-free electrodynamics. Within this framework, the 
plasma in the magnetosphere above the accretion disk is assumed to 
have such a low density that it is completely unimportant dynamically. 
At the same time, the plasma is 
dense enough to carry the necessary currents and charges without 
significant dissipation. This framework has proven to be very 
useful as it apparently provides the minimal nontrivial level 
of description required by the magnetospheric conditions. 
Under the usual additional assumptions of time stationarity 
and axisymmetry, the main fundamental mathematical formulation
of this framework is the so-called Grad--Shafranov equation. 
Over the years, there have been a number of attempts to solve 
this rather nontrivial nonlinear Partial Differential Equation 
(PDE) in the context of a black-hole magnetosphere with open magnetic
field. These studies include both semi-analytical models that 
use some sort of a self-similar ansatz (e.g., BZ77), and also 
the most general numerical computations (e.g., Macdonald~1984; 
Fendt~1997; Komissarov~2001). At the same time, however, there 
have been, to the best of my knowledge, no numerical or analytical 
attempts to solve the Grad--Shafranov equation in the context of 
the magnetically-linked black hole--disk system.

The goal of this paper is to remedy this situation by providing the first
numerical solution of the Grad--Shafranov equation for the MC configuration.
In order to achieve this goal, one first needs to examine the structure 
of the equation and, in particular, understand the role of, and devise 
a proper mathematical treatment for, the singular surfaces of the 
Grad--Shafranov equation, namely the Event Horizon and the Light 
Cylinder. One very important thing I would like to emphasize in 
this regard is that the condition of regularity at the light cylinder 
is crucial; indeed, it is this conditions that enables one to fix the 
poloidal current function and hence the toroidal magnetic field.
This point of view is very close in spirit to that of Beskin \& 
Kuznetsova (2000), who suggested a similar approach for the full-MHD 
case. I also would like to add that, in this respect, the situation is 
very similar to the problem of axisymmetric pulsar magnetosphere, where 
the light-cylinder regularity condition plays a similar role (Contopoulos 
et al.~1999; Uzdensky~2003). As for the event horizon, it plays only a 
passive role here, similar to that of the asymptotic infinity (see, e.g., 
Punsly~1989; Punsly \& Coroniti~1990). Thus, the horizon is, in a sense,
less important; for example, one cannot set any boundary conditions on it
(e.g., Beskin~1997; Beskin \& Kuznetsova 2000).

Although the most interesting and general case is that of a 
rapidly-rotating Kerr black hole, in the present work I restrict 
myself to the simpler case of a nonrotating, Schwarzschild black 
hole. This work should thus be viewed as a first starting step. 
Even the Schwarzschild case, however, is not entirely trivial, 
because the disk and, hence, the magnetosphere are still 
(nonuniformly!) rotating, and thus the Grad--Shafranov equation 
is still nonlinear and has singular surfaces. 

Considering the Schwarzschild case first has a purely technical 
advantage of having to deal with fewer terms in the equations.
In addition, a closed-field solution is certain to exist in 
this case; whether it exists in the more general Kerr case 
is not so clear. Indeed, it may be that, for a sufficiently 
rapidly-rotating hole, a completely closed (i.e., with all 
the field lines threading both the event horizon and the disk) 
configuration may not be possible, that is, some fraction of 
the field lines may have to be open and to extend from the hole 
to infinity. This scenario could be characterized as a hybrid between 
the BZ77 and~MC configurations, as suggested by Wang~et~al. (2002, 2003). 
I plan to consider such a configuration in full Kerr geometry in the near 
future.

Finally, I would like to remark that a black hole--disk MC configuration 
differs greatly from the case where the central object is a star 
with a highly-conducting surface, such as a neutron star or a young 
star. In that latter case, the differential rotation between the disk 
and the conducting star inevitably leads to the inflation and opening 
of the field lines on the rotation time scale. This, in turn, makes a 
steady state impossible (e.g., van~Ballegooijen 1994; Lovelace~et~al.
1995; Uzdensky~et~al. 2002). In contrast, in the case of a black 
hole being the central object, a steady configuration is, in principle, 
possible (at least in the Schwarzschild case). This is because the 
rather large ``effective resistivity'' of the event horizon (in the 
Membrane-Paradigm description; see Znajek 1977, 1978; Damour~1978;
Thorne~et~al. 1986) makes it possible for the field lines rotating 
with the disk's angular velocity to slip through the horizon. This 
important fact makes the study of a black hole's magnetosphere conceptually 
simpler than that of a regular star, even though the proper treatment 
of the black-hole case is unavoidably plagued with technical difficulties, 
such as having to work in curved space-time.

In \S~\ref{sec-model} I outline the basic equations that describe
a stationary axisymmetric force-free magnetosphere in Schwarzschild
geometry; in particular, I discuss the Grad--Shafranov Equation.
In the same section I also describe the boundary conditions 
pertinent to the MC configuration under consideration.
In \S~\ref{sec-regularity} I discuss the singular surfaces of
the Grad--Shafranov equation and the regularity conditions set
on these surfaces. In particular, \S~\ref{subsec-EH} is devoted 
to the regularity condition at the event horizon and \S~\ref
{subsec-LC} is devoted to the light-cylinder regularity condition. 
Next, in \S~\ref{sec-zero-rotation}, I consider a particularly
simple but very important case of a nonrotating disk around a 
nonrotating black hole. In this case the light cylinder merges 
with the horizon and the Grad--Shafranov becomes linear. I solve 
this equation numerically and find that the radial magnetic field 
is uniform on the event horizon. In \S~\ref{sec-slow} I consider 
the limit of a slowly-rotating disk and derive explicit analytical 
expressions for the location and the shape of the light cylinder 
and for the poloidal current. I illustrate these ideas by considering 
one particular manifestation of the slow-rotation limit, namely, 
a Keplerian disk. I present my numerical solution of the full, 
nonlinear Grad--Shafranov equation for this case and find a perfect 
agreement between the numerical results and the above-mentioned 
analytical predictions. Finally, in \S~\ref{sec-conclusions} I 
present my conclusions and discuss possible extensions of my
present work and the directions for future research.


\section{Stationary Axisymmetric Force-Free Magnetosphere in 
Schwarzschild Geometry}
\label{sec-model}

I consider a steady-state force-free magnetosphere 
of a Schwarzschild Black Hole surrounded by a thin,%
\footnote
{A more realistic case of a thick disk or torus 
requires a much more sophisticated physical model, 
because in this case there may not be such a clear-cut 
distinction between a dense disk and a tenuous force-free 
corona; in addition, the magnetic field is likely to be 
very intermittent and nonstationary.}
rotating, infinitely-conducting accretion disk. I am interested 
only in the large-scale magnetic field, ignoring any small-scale 
and intermittent field structures. I assume the topology of this 
global magnetic field to be such that all the field lines connect 
the disk to the event horizon (or, rather, the stretched horizon 
of the Membrane Paradigm, see Thorne~et~al. 1986) of the black hole 
(see Fig.~\ref{fig-geometry}); in particular this means that there 
are no open field lines extending out to infinity. I also assume 
that the system possesses axial symmetry and reflection symmetry 
with respect to the equatorial disk plane. 

Magnetic field lines are assumed to be frozen into the disk. 
Also, whereas inside the disk the magnetic field is considered 
to be dynamically unimportant, in the very tenuous magnetosphere 
above the disk the electromagnetic forces are assumed to be dominant 
over the inertial, gravitational, and pressure forces. Under these 
conditions, the structure of the magnetosphere is governed by the 
force-free equation:
\begin{equation} 
\rho_e {\bf E} \, + {{\bf j\times B}\over c} = 0 \, ,
\label{eq-force-free}
\end{equation}
where all the electromagnetic quantities are measured by the
Fiducial Observers (FIDOs), see MT82 and Thorne~et~al. (1986).

At this point, however, I have to make the following remark.
In a realistic situation the force-free approximation is bound 
to break down close enough to the event horizon, as the plasma 
inertia and gravity necessarily start to dominate the dynamics
there. In particular, the black hole is always enveloped by a fast
magnetosonic critical surface. Nevertheless, as the plasma density 
is taken to zero, this fast critical surface moves in infinitesimally 
close to the horizon (e.g., Beskin~1997). This fact makes it possible 
to effectively extend the domain of validity of the force-free 
approximation all the way to the horizon. At the same time, however,
this ``full-MHD origin'' of the force-free equation is important,
because it provides one with an additional condition that leads to 
a way to regularize the force-free solution at the event horizon
(see \S~\ref{subsec-EH}). Such a regularization is possible because
the solutions of the full-MHD set of equations are automatically 
free of certain singularities that are admitted, in principle, by 
the force-free equation.

The geometry of space-time around a Schwarzschild Black hole is 
described by
\begin{equation}
ds^2 = - \alpha^2 dt^2 + \alpha^{-2} dr^2 + 
r^2 d\theta^2 + r^2 \sin^2\theta \, d\phi^2 \, ,
\label{eq-metric}
\end{equation}
where 
\begin{equation}
\alpha(r) \equiv \sqrt{1-{r_s\over r}} 
\label{eq-alpha}
\end{equation}
is the lapse function and $r_s \equiv 2GM/c^2$ is the Schwarzschild 
radius of the event horizon. 

I employ the 3+1 split of laws of electrodynamics, introduced by~MT82. 
In this approach, Maxwell's equations in Schwarzschild geometry take 
the following form:
\begin{eqnarray}
&&\nabla\cdot{\bf E} = 4 \pi \rho_e \, ,     \label{eq-Maxwell-1} \\
&&\nabla\cdot{\bf B} = 0 \, ,                \label{eq-Maxwell-2} \\
&&\nabla\times(\alpha{\bf E}) = 
- {1\over c}\, \partial_t {\bf B} \, ,       \label{eq-Maxwell-3} \\
&&\nabla\times(\alpha{\bf B}) = {1\over c}\,  \partial_t {\bf E} +
{{4\pi\alpha}\over c}\,  {\bf j} \, .        \label{eq-Maxwell-4} 
\end{eqnarray}

An axisymmetric magnetic field can be written in terms of two functions
$\Psi(r,\theta)$ and $I(r,\theta)$ as
\begin{equation}
{\bf B(r,\theta)} = {\bf B}_{\rm pol} + {\bf B}_{\rm tor} =
\nabla\Psi \times \nabla\phi +{I\over{\alpha}}\, \nabla\phi \, ,
\label{eq-B}
\end{equation}
where $\Psi$ is the poloidal magnetic flux function and $I$ 
is $(2/c)$ times the poloidal electric current flowing through
the circular loop $r={\rm const}$, $\theta={\rm const}$. Note 
that the choice of sign in this definition of~$I$ is consistent 
with the choice made by Mobarry \& Lovelace (1986) but is opposite 
to that by~MT82.

I use the orthonormal basis $\{ {\bf e}_{\hat{\alpha}}\}= 
\{ {\bf e}_{\hat{r}},{\bf e}_{\hat{\theta}},{\bf e}_{\hat{\phi}}\}$,
where ${\bf e}_{\hat{\alpha}} = g_{\alpha\alpha}^{-1/2} \partial_{\alpha}$
(note: there is no summation over $\alpha$ in this expression), that is
\begin{equation}
{\bf e}_{\hat{r}} = \alpha \, \partial_r, \quad
{\bf e}_{\hat{\theta}} = {1\over r}  \, \partial_\theta, \quad
{\bf e}_{\hat{\phi}} = {1\over{r\sin\theta}} \, \partial_\phi\, .
\label{eq-basis}
\end{equation}

In Schwarzschild metric~(\ref{eq-metric}), the 3-gradient of an 
arbitrary function $\chi(r,\theta,\phi)$ is calculated as follows:
\begin{equation}
\nabla\chi = \alpha \, (\partial_r \chi) \, {\bf e}_{\hat{r}} +
{1\over r} \, (\partial_\theta \chi) \, {\bf e}_{\hat{\theta}} +
{1\over{r\sin\theta}} \, (\partial_\phi \chi) \, {\bf e}_{\hat{\phi}}\, .
\label{eq-grad-Schwarzschild}
\end{equation}
Thus,
\begin{equation}
\nabla\Psi(r,\theta) = \alpha \, (\partial_r \Psi) \, {\bf e}_{\hat{r}}+
{1\over r} \, (\partial_\theta \Psi) \, {\bf e}_{\hat{\theta}} \, ,
\label{eq-grad-Psi}
\end{equation}
and
\begin{equation}
\nabla\phi={1\over{r\sin\theta}} \, {\bf e}_{\hat{\phi}}\, .
\label{eq-grad-phi}
\end{equation}

Applying this to expression~(\ref{eq-B}) for the magnetic field 
components, one gets
\begin{equation}
{\bf B}_{\rm pol} = \nabla\Psi\times \nabla\phi =
{1\over{r\sin\theta}} \, {{\partial\Psi}\over{r\partial\theta}} \, 
{\bf e}_{\hat{r}}-
{\alpha\over{r\sin\theta}} \, {{\partial\Psi}\over{\partial r}} \, 
{\bf e}_{\hat{\theta}} \, ,
\label{eq-B_pol}
\end{equation}
and
\begin{equation}
{\bf B_{\rm tor}} = {I\over\alpha}\nabla\phi =
{I\over{\alpha r\sin\theta}} \, {\bf e}_{\hat{\phi}} \, .
\label{eq-B_tor}
\end{equation}

As for the electric field, it can be shown that, under the conditions 
of axisymmetry, stationarity, and ideal magnetohydrodynamics (MHD), i.e., 
${\bf E\cdot B}=0$, it is given by 
\begin{eqnarray}
E_{\hat{\phi}} &=& 0 \, , \label{eq-Ephi=0}   \\ 
{\bf E}_{\rm pol} &=& -\, {1\over{\alpha c}}\ \Omega(\Psi)\, \nabla\Psi \, ,
\label{eq-E}
\end{eqnarray}
where $\Omega(\Psi)$ is the angular velocity of the field lines.

As I have mentioned earlier, the main equation determining the structure 
of the magnetosphere is the force-free equation~(\ref{eq-force-free}).
Upon examining the toroidal component of this equation, one immediately
sees that poloidal current~$I$ has to be constant along magnetic field 
lines:
\begin{equation}
I = I(\Psi) \, .
\label{eq-I=I_of_Psi}
\end{equation}

Thus, the electromagnetic field in a stationary axisymmetric 
ideal-MHD force-free magnetosphere is completely described by
the poloidal magnetic flux function $\Psi(r,\theta)$ and two 
functions of $\Psi$, namely, $\Omega(\Psi)$ and~$I(\Psi)$.

Next, one can show that the two poloidal components of the force-free 
equation~(\ref{eq-force-free}) can be combined into the force-free 
Grad--Shafranov equation for the poloidal flux function $\Psi(r,\theta)$.
In the most general Kerr-metric case this equation has first been derived 
in the full four-dimensional framework by BZ77 and then later by MT82, who 
had used the language of the 3+1 split (see also Beskin~1997). Subsequently,
it has been generalized to full general-relativistic MHD including the 
effects of plasma inertia and pressure (e.g., Nitta~et~al. 1991; Beskin 
\& Par'ev 1993). In the specific case of Schwarzschild geometry, 
the full-MHD Grad--Shafranov equation, along with its simpler, force-free 
version, has been derived and discussed by Mobarry \& Lovelace (1986). 
Here I shall use their force-free Grad--Shafranov equation (44); in 
my notation it can be written as follows (I set the speed of light~$c$ 
and the gravitational constant $G$ both equal to one throughout this paper):
\begin{equation} 
(\alpha^2 - R^2 \Omega^2)\,  \Delta^* \Psi =
{\alpha^4\over{2R^2}} \, \nabla\Psi \cdot 
\nabla\biggl({{R^4\Omega^2}\over\alpha^4} \biggr) - II'(\Psi)\, ,
\label{eq-GS-ML}
\end{equation}
where
\begin{equation} 
R \equiv r \sin\theta \, ,
\label{R-def}
\end{equation}
and 
\begin{equation} 
\Delta^* \Psi \equiv \partial_r (\alpha^2 \partial_r \Psi) +
{\sin\theta\over r^2} \, \partial_\theta \,  
\biggl({1\over{\sin\theta}}\, \partial_\theta\Psi \biggr)
\label{eq-GS-operator}
\end{equation}
is the linear Grad--Shafranov operator in Schwarzschild geometry.

Equation~(\ref{eq-GS-ML}) is a nonlinear second-order elliptic 
PDE for the function $\Psi(r,\theta)$; the right-hand side (RHS) 
of this equation contains no second-order derivatives of~$\Psi$, 
but does contain terms that are linear and quadratic in the first 
order derivatives of~$\Psi(r,\theta)$. One can rewrite this equation 
in the following alternative form, more suitable for further 
calculations:
\begin{eqnarray}
(\alpha^2-R^2\Omega^2)\,  \biggl(\alpha^2\partial_r^2\Psi+
{1\over r^2}\, \Delta_{\theta}^*\Psi\biggr) &+&
\alpha^2 [\alpha^2(r)]' \, \partial_r\Psi =                   \nonumber \\
\Omega\Omega'(\Psi)\, R^2 |\nabla\Psi|^2 \, &-& II'(\Psi)+ 
\Omega^2 \biggl( 2\alpha^2 \sin^2\theta \, r\, \partial_r\Psi+
\sin{2\theta}\, \partial_\theta\Psi \biggl) \, ,
\label{eq-GS-1}
\end{eqnarray}
where
\begin{equation} 
\Delta_{\theta}^*\Psi \equiv \sin\theta\, {\partial\over{\partial\theta}}\,
\biggl( {1\over{\sin\theta}} \,{\partial\Psi\over{\partial\theta}} \biggr)
\label{eq-GS-operator_theta}
\end{equation}
is the poloidal-angle part of the Grad--Shafranov operator and
\begin{equation}
|\nabla\Psi|^2=\alpha^2\biggl({{\partial\Psi}\over{\partial r}}\biggr)^2+
{1\over{r^2}}\, \biggl({{\partial\Psi}\over{\partial\theta}}\biggr)^2\, .
\label{eq-gradPsi-squared}
\end{equation}

In flat space with Euclidean geometry, $\alpha=1={\rm const}$, 
equation (\ref{eq-GS-1}) becomes
\begin{eqnarray}
(1-R^2\Omega^2)\, \biggl(\partial_r^2\Psi+
{1\over r^2}\, \partial_{\theta}^2\Psi\biggr) &-&
(1+R^2\Omega^2)\, {{\cot\theta}\over{r^2}}\, \partial_\theta \Psi -
2\Omega^2 \, r\sin^2\theta \, \partial_r \Psi              \nonumber \\
&=& \Omega\Omega'(\Psi)\, R^2\, |\nabla\Psi|^2  - II'(\Psi) \, ,
\label{eq-GS-Euclidean}
\end{eqnarray}
which can be immediately recognized as the familiar pulsar equation 
with $\Omega=\Omega(\Psi)\neq
{\rm const}$ (e.g., Okamoto~1974).

In Schwarzschild geometry, the radial derivative of the square of 
the lapse function $\alpha$ is $[\alpha^2(r)]'=r_s/r^2$, and so 
one can rewrite~(\ref{eq-GS-1}) in the following final form:
\begin{eqnarray}
(\alpha^2-R^2\Omega^2) \, \biggl(\alpha^2 \partial_r^2\Psi+
{1\over r^2}\, \partial_{\theta}^2\Psi\biggr) &-&
(\alpha^2+R^2\Omega^2)\, {{\cot\theta}\over{r^2}}\, \partial_\theta \Psi +
{\alpha^2\over r}\, \biggl({r_s\over r}-2R^2 \Omega^2 \biggr)\, 
\partial_r\Psi  \nonumber \\
&=& \Omega\Omega'(\Psi)\, R^2\, |\nabla\Psi|^2  - II'(\Psi) \, .
\label{eq-GS-2}
\end{eqnarray}

Let us now discuss the boundary conditions that supplement 
equation~(\ref{eq-GS-ML}) [or its equivalent forms~(\ref{eq-GS-1}) 
and~(\ref{eq-GS-2})]. First, the rotation axis is a field line, so 
let us set
\begin{equation}
\Psi(r,\theta=0) = 0 \, .
\label{eq-bc-axis}
\end{equation}

Second, since we are considering the case where all the field 
lines are closed, i.e., go from the black hole to the disk,
the flux function has to vanish at infinity:
\begin{equation}
\Psi(r=\infty,\theta)=0\, .
\label{eq-bc-infinity}
\end{equation}

Third, we consider a disk that does not extend all the way to the 
event horizon but instead ends abruptly at a certain inner-edge 
radius~$r_{\rm in}$. The most natural choice for~$r_{\rm in}$ is
$r_{\rm ISCO}$, the radius the Innermost Stable Circular Orbit 
(ISCO) (equal to~$6M$ for a Schwarzschild black hole); at this 
point in the discussion, however, one can just leave~$r_{\rm in}>r_s$ 
unspecified. I assume that the magnetic field lines are frozen into 
the disk beyond~$r_{\rm in}$, with some given (but, at this point, 
arbitrary) magnetic flux distribution:
\begin{equation}
\Psi(r\geq r_{\rm in}, \theta=\pi/2) = \Psi_d(r) \, .
\label{eq-bc-disk}
\end{equation}

The region $(r_s\leq r\leq r_{\rm in},\theta=\pi/2)$ that separates 
the disk from the black hole is generally referred to as the plunging 
region. Here the infalling matter can no longer be supported against 
gravity by the centrifugal force;  hence, it presumably falls very 
rapidly towards the black hole. As a result, magnetic loops are 
strongly stretched in the radial and azimuthal directions and the 
vertical magnetic field is greatly diminished.%
\footnote
{Thus, I here assume that the magnetic support of matter against
gravity in the plunging region is also insufficient (e.g., Li~2003). 
A possible alternative scenario would be the one with a force-free 
gap between the disk and the event horizon, whereby the magnetic 
field in the gap is strong enough to prevent the matter from falling 
onto the black hole. The boundary condition in this case would be the 
requirement that the magnetic field be perpendicular to the equator, 
i.e., $\partial\Psi/\partial\theta=0$ for $r_s\leq r\leq r_{\rm in}$, 
$\theta=\pi/2$.}
Under these circumstances, it is natural to choose the plunging-region 
boundary condition in the following form:
\begin{equation}
\Psi(r_s\leq r\leq r_{\rm in},\theta=\pi/2)=\Psi_{\rm max}\equiv 
\Psi_d(r_{\rm in}) = {\rm const} \, .
\label{eq-bc-gap}
\end{equation}

Finally, one cannot and, indeed, need not specify any boundary 
conditions at the event horizon. This is because the horizon, 
being an analog of spatial infinity, cannot emit waves that 
carry information outward and is thus casually disconnected 
from the outside magnetosphere (e.g., Punsly~1989; Punsly \& 
Coroniti 1990). Within the force-free framework, the mathematical 
basis for this statement lies in the fact that the horizon is a 
singular surface of the Grad--Shafranov equation. Therefore, one 
can only apply some {\it regularity conditions} there (see \S~\ref
{subsec-EH} for a detailed discussion).

Notice that specifying the boundary conditions is not enough for 
a complete problem set-up. Indeed, the force-free Grad--Shafranov
equation~(\ref{eq-GS-ML}) [and its equivalent forms~(\ref{eq-GS-1})
and~(\ref{eq-GS-2})] also involves two, a priori unknown, functions
of~$\Psi$, i.e., $\Omega(\Psi)$ and~$I(\Psi)$; therefore, one also
needs to discuss what determines these two functions. First, in the
problem considered in this paper, it is in fact very easy to find 
$\Omega(\Psi)$. Indeed, since the field lines are assumed to be frozen 
into the disk, each field line has to rotate with the angular velocity 
of its footpoint on the disk surface:
\begin{equation}
\Omega(\Psi) = \Omega_d[r_0(\Psi)] \, .
\label{eq-Omega}
\end{equation}
Here, $\Omega_d(r)$ is a prescribed disk rotation law (e.g., 
Keplerian) and $r_0(\Psi)$ is the radial position of the 
footpoint of field line~$\Psi$ on the surface of the disk; 
it is related to the function $\Psi_d(r)$ via $\Psi_d[r_0(\Psi)]= 
\Psi$.

As for the other function, i.e., $I(\Psi)$, it is not physically 
appropriate to try to specify it explicitly at the surface of the 
disk or at the event horizon. The basic reason for this is the 
following. The Grad--Shafranov equation is a second-order PDE with 
two singular surfaces and two unknown functions of~$\Psi$ (i.e., 
two ``integrals of motion''), $\Omega(\Psi)$ and~$I(\Psi)$. 
Hence, one can (and indeed needs to) specify only {\it two} 
functions at the boundaries (see Beskin~1997) and I have 
already chosen both of these functions, $\Psi_d(r)$ and 
$\Omega(\Psi)=\Omega_K[r_0(\Psi)]$, to be specified on 
the disk surface [see eqs.~(\ref{eq-bc-disk}) and~(\ref
{eq-Omega})]. Therefore, trying to specify directly one 
more function, such as~$I(\Psi)$, would over-constrain 
the system. Instead, as I discuss in detail in \S~\ref{subsec-LC}, 
the correct way to determine $I(\Psi)$ is through the use of the 
regularity condition at the inner light cylinder.


\section{Regularity Conditions on the Singular Surfaces 
of the Grad--Shafranov Equation (for $\Omega\neq 0$)}
\label{sec-regularity}

As one can easily see from the Grad--Shafranov equation 
in the form~(\ref{eq-GS-2}), a force-free magnetosphere 
of a Schwarzschild black hole (with the closed-field 
configuration considered here) has two regular singular 
surfaces on which the coefficients in front of the highest 
order derivatives of~$\Psi$ vanish. These are the {\it Event 
Horizon} $\alpha=0$ (or $r=r_s=2M$) and the {\it Light Cylinder} 
$\alpha^2=R^2\Omega^2$. On each of these two surfaces I am going 
to impose a corresponding {\it regularity condition}; these two 
regularity conditions are going to be indispensable in my analysis.

The first of the two surfaces, i.e., the event horizon, is singular 
because the coefficient $\alpha^2\equiv 1-r_s/r$ in front of the 
second-order radial (i.e., normal to this surface) derivative 
becomes zero. This surface is the limit of the fast magnetosonic 
surface as the plasma density goes to zero. Correspondingly, as 
one analyses the low-density limit of the full-MHD Grad--Shafranov 
equation, one finds that the critical condition on the fast surface 
turns into the event-horizon regularity condition (Beskin 1997). 
Since one of the boundaries of the domain coincides with the event 
horizon, I shall use the regularity condition at the horizon in 
lieu of a boundary condition for $\Psi(r,\theta)$ at $r=r_s$ 
(see \S~\ref{subsec-EH}).

The second singular surface, namely, the light cylinder,
can be characterized as the surface where the magnitude 
of the electric field is equal to that of the poloidal 
magnetic field, and, correspondingly, where the toroidal 
velocity of the rotating field lines, $v_{\bf B} =R\Omega/\alpha$ 
equals the speed of light.%
\footnote
{I follow here the terminology used by Beskin (1997) and 
reserve the term ``light cylinder'' to denote the surface 
where $|{\bf E}|=|{\bf B}_{\rm pol}|$ and the term ``light
surface'' to denote the surface where $|{\bf E}|=|{\bf B}|$. 
In our problem, there is only one (inner) light cylinder,
corresponding to the Alfv{\'e}n surface in the limit of
zero plasma density; the light surface coincides with the 
event horizon.} 
I shall use the regularity condition at the light cylinder as the main 
condition that fixes the function~$I(\Psi)$ (see \S~\ref{subsec-LC}).

It is important to note that, in contrast to a {\it rigidly-rotating} 
pulsar magnetosphere, in the case of an accreting black hole, where 
the field lines are tied to a {\it differentially-rotating} Keplerian 
disk, the angular velocity $\Omega$ is a (non-constant) function 
of~$\Psi$. Hence, the spatial location and even the shape of the 
light cylinder are not known a priori; they need to be determined 
self-consistently as a part of the whole solution. Also note that 
the light cylinder, that one has to deal with in the problem considered
here, is the so-called inner light cylinder. In a more general situation, 
where some of the field lines are open and extend to infinity, there 
may be two light cylinders: the outer one and the inner one (Znajek~1977; 
BZ77). The outer light cylinder is crossed at large distances by the open 
field lines and is a direct analog of the light cylinder in pulsar 
magnetospheres. The inner light cylinder is crossed much closer to 
the black hole by the field lines threading the event horizon; its 
existence is a purely general-relativistic (GR) effect; it is due 
to the fact that the expression for the electric field~(\ref{eq-E}) 
has the additional factor $1/\alpha$ as compared with the non-GR 
expression.


\subsection{Regularity Condition at the Event Horizon}
\label{subsec-EH}

Let us now discuss the behavior of the flux function in the vicinity 
of the event horizon. In general, equation~(\ref{eq-GS-2}) admits 
solutions that are not regular at the horizon. In particular, the 
two indicial exponents of the linear part of equation~(\ref{eq-GS-2}) 
are equal to zero and one; hence, one can expect the asymptotic expansion 
near $r=r_s$ to contain logarithmic terms in addition to pure power laws
(e.g., Bender \& Orszag 1978). Taking into account that the magnetic flux 
must be finite on the horizon (see below), one can write the first few 
terms in the asymptotic expansion of a general solution of the full 
nonlinear equation~(\ref{eq-GS-2}) as
\begin{eqnarray}
\Psi(r\rightarrow r_s,\theta) &=& \Psi_0(\theta)+
\tilde{\Psi}_1(\theta)(r-r_s)\log(r-r_s) +\Psi_1(\theta)(r-r_s) \nonumber \\
&+& \hat{\Psi}_2 (\theta) (r-r_s)^2 \log^2(r-r_s)+ 
\tilde{\Psi}_2 (\theta) (r-r_s)^2 \log(r-r_s) \nonumber \\
&+& \Psi_2(\theta)(r-r_s)^2 \ + \ ... 
\label{eq-EH-expansion}
\end{eqnarray}
Here, the presence of the $(r-r_s)^2 \log^2(r-r_s)$-term is due
to the nonlinear term $R^2\,\Omega\Omega' |\nabla\Psi|^2$
that appears on the RHS of the Grad--Shafranov equation.

A physically-reasonable solution, however, should be free of the
logarithmic terms, as we shall see below. One can trace the origin 
of this regularity requirement to the full-MHD framework's condition 
of regularity at the fast magnetosonic surface, as one makes the 
transition to the force-free limit (e.g., Beskin 1997; Beskin \& 
Kuznetsova 2000). The physical meaning of the event-horizon regularity 
condition has been discussed beautifully by MT82 and by Thorne et al. 
(1986). My discussion here follows their line of thought. 

According to MT82 and Thorne~et~al. (1986), the only condition 
that one can impose at the event horizon is the requirement that 
the physically-reasonable Freely-Falling Observer (FFO) measures 
{\it finite} fields ${\bf E}_{FFO}$ and ${\bf B}_{FFO}$ near the 
horizon. In order to determine the behavior of the fields measured 
by the FIDO, one then has to perform a Lorentz transformation from 
the FFO frame to the FIDO frame. This transformation is a radial 
boost with the velocity $v=2M/r$, which corresponds to $\gamma=1/\alpha$. 
The result is:
\begin{equation}
{\bf E}_\perp , \ {\bf B}_\perp \qquad - \quad {\rm finite;} 
\label{eq-EH-1}
\end{equation}
\begin{equation}
{\bf E}_\parallel , \ {\bf B}_\parallel \propto {1\over\alpha},
\quad \alpha\rightarrow 0 \, ;
\label{eq-EH-2}
\end{equation}
but
\begin{equation}
{\bf E}_\parallel - {\bf n}\times {\bf B}_\parallel \sim \alpha, 
\quad \alpha \rightarrow 0 \, .
\label{eq-EH-3}
\end{equation}
(Here the $\perp$ sign describes the perpendicular to the event horizon, 
i.e., radial, component, and the $\parallel$ sign describes the parallel 
to the event horizon, i.e., the $\phi$ and $\theta$, components.)

First, from equation~(\ref{eq-EH-1}) it follows that $\Psi(r_s,\theta)$ 
must be finite, as we have already anticipated.

Next, consider equation~(\ref{eq-EH-3}); it has two components 
($\phi$ and~$\theta$) and I shall consider them separately. 
First, since $E_{\hat{\phi}}=0$ according to equation~(\ref
{eq-Ephi=0}), the $\phi$-component of equation~(\ref{eq-EH-3})
gives
\begin{equation} 
B_{\hat{\theta}} = O(\alpha) \qquad \Rightarrow \qquad 
{\partial\Psi\over{\partial r}} = \ {\rm finite \ at} \quad \alpha=0\, .
\label{eq-EH-Btheta=0}
\end{equation}
Thus, $\tilde{\Psi}_1(\theta)\equiv 0$ and the expansion~(\ref
{eq-EH-expansion}) near the horizon becomes:
\begin{eqnarray}
\Psi(r,\theta) &=& \Psi_0(\theta) + \Psi_1(\theta) (r-r_s) + 
\hat{\Psi}_2(\theta) (r-r_s)^2\log^2(r-r_s) \nonumber \\
&+& \tilde{\Psi}_2(\theta)(r-r_s)^2\log(r-r_s) \ +
\ \Psi_2(\theta)(r-r_s)^2 \ + \ ...
\label{eq-EH-Psi}
\end{eqnarray}

The condition~(\ref{eq-EH-Btheta=0}) may seem trivial, 
but notice that everywhere in equation~(\ref{eq-GS-2}) 
the first- and second-order radial derivatives of $\Psi$ 
appear only in a combination with~$\alpha$ or~$\alpha^2$. 
Thus, without condition~(\ref{eq-EH-Btheta=0}), it would 
in principle be possible for these derivatives to diverge 
in the limit~$r\rightarrow r_s$, while leaving the product 
$\alpha^2\partial_r^2 \Psi$ finite and comparable with all 
the other (regular) terms near the event horizon. It is only 
due to condition~(\ref{eq-EH-Btheta=0}) that one can discard 
all such solutions and hence drop all the terms proportional 
to $\alpha$ when applying equation~(\ref{eq-GS-2}) at the 
horizon. One then gets the following equation:
\begin{equation}
II'(\Psi_0) = \Omega(\Psi_0) \sin\theta
\biggl[ \Omega'\sin\theta \biggl({{d\Psi_0}\over{d\theta}}\biggr)^2+
\Omega \cos\theta {{d\Psi_0}\over{d\theta}} +
\Omega \sin\theta {{d^2\Psi_0}\over{d\theta^2}} \biggr] \, .
\label{eq-EH-II'}
\end{equation}
In other words, equation~(\ref{eq-EH-II'}) is obtained 
by substituting expansion~(\ref{eq-EH-Psi}) into the 
Grad--Shafranov equation~(\ref{eq-GS-2}) and looking 
at the terms of lowest order in $(r-r_s)$. Furthermore, 
when one looks at the balance in the next two orders 
in $(r-r_s)$, namely $O[(r-r_s)\log^2(r-r_s)]$ and 
$O[(r-r_s)\log(r-r_s)]$, one can deduce that 
$\hat{\Psi}_2(\theta)=0=\tilde{\Psi}_2(\theta)$, 
and so $\partial_r^2\Psi$ must be finite at the 
event horizon.

It is interesting to note that not only the second, but also the 
first radial derivatives of $\Psi$ drop out of the Grad--Shafranov 
equation when $\alpha$ is set to zero. One thus sees that, upon using 
the regularity condition~(\ref{eq-EH-3}), the Grad--Shafranov {\it 
differential} equation becomes an {\it algebraic} equation at the 
event horizon, as far as the radial direction is concerned; it is 
of course still a differential equation in the $\theta$-direction. 
The resulting equation~(\ref{eq-EH-II'}) can be viewed as an 
Ordinary Differential Equation (ODE) that determines the function 
$\Psi_0(\theta)\equiv\Psi(r=r_s,\theta)$, provided that both 
$I(\Psi)$ and $\Omega(\Psi)$ are known. 

Equation~(\ref{eq-EH-II'}) can be readily integrated, resulting in
\begin{equation}
I(\Psi_0) = \pm \Omega(\Psi_0) \sin\theta\, {{d\Psi_0}\over{d\theta}}\, ,
\label{eq-EH-I-1}
\end{equation}
where I set the integration constant equal to zero because
I require the poloidal current $I(\Psi)$ to vanish at the 
pole $\theta=0$. One can integrate~(\ref{eq-EH-I-1}) further
to determine (albeit implicitly) the function $\Psi_0(\theta)$:
\begin{equation}
\int\limits_0^{\Psi_0} {{\Omega(\Psi_0)}\over{\pm I(\Psi_0)}} d\Psi_0 =
\int\limits_0^{\theta} {d\theta\over{\sin\theta}}\, ,
\label{eq-Psi0-implicit}
\end{equation}
where I have used the boundary condition $\Psi_0(\theta=0)=0$.

Notice that, in principle, the regularity condition~(\ref{eq-EH-I-1}), 
derived from the $\phi$~component of equation~(\ref{eq-EH-3}), admits
both plus and minus signs. In order to break this degeneracy, one needs 
to use the remaining condition, namely the  $\theta$ component of 
equation~(\ref{eq-EH-3}):
\begin{equation} 
E_{\hat{\theta}}+B_{\hat{\phi}}=O(\alpha), \qquad \alpha \rightarrow 0\, .
\label{eq-EH-3_theta}
\end{equation}
According to equations~(\ref{eq-B_tor}) and~(\ref{eq-E}), both 
$B_{\hat{\phi}}$ and~$E_{\hat{\theta}}$ are of order $1/\alpha$ 
near the event horizon, which is consistent with equation~(\ref{eq-EH-2}).
Condition~(\ref{eq-EH-3_theta}), however, tells us that these fields 
balance each other out to the lowest order in~$\alpha$. Using 
expressions~(\ref{eq-E}) for~${\bf E}$ and~(\ref{eq-B_tor}) 
for~$B_{\hat{\phi}}$, one immediately obtains:
\begin{equation}
I(\Psi)=+\Omega(\Psi) \sin\theta \, {{d\Psi_0(\theta)}\over{d\theta}} \, .
\label{eq-EH-I-2}
\end{equation}
This equation is identical to equation~(\ref{eq-EH-I-1}) with the plus 
sign; thus, condition~(\ref{eq-EH-3}), derived from the requirement that 
the FFO falling onto the black hole (as opposed to the one coming out of 
it) measures finite fields, enables one to break the sign degeneracy in 
equation~(\ref{eq-EH-I-1}).

Condition~(\ref{eq-EH-I-2}) has been derived in the general case 
of a Kerr black hole by Znajek (1977). Note that the corresponding 
equation (7.29b) of MT82, when applied to the nonrotating black hole 
case~$\Omega_H=0$, has the opposite sign (minus). This is because 
they define $I(\Psi)$ with the minus sign [e.g., their equation (4.8)] 
as compared with my equation~({\ref{eq-B_tor}). Also note that Mobarry 
\& Lovelace (1986) use the same sign convention for~$I$ as is used in
this paper, but they still cite the MT82 condition with the minus sign.

To sum up, one can impose a certain regularity condition at the 
event horizon; this condition is the rudiment of the full-MHD 
fast magnetosonic critical condition in the limit of vanishing 
plasma density (see Beskin~1997). In the case $\Omega\neq 0$, 
the event-horizon regularity condition can be viewed as the 
equation that determines the values of~$\Psi$ on the horizon; 
thus it is in a sense similar to prescribing a Dirichlet-type 
boundary condition at the event horizon. The reason for this is 
that not only the second-, but also the first-order radial derivatives 
drop out of this equation when $\alpha=0$. It is interesting to note 
that, in contrast, in the case of a nonrotating field [$\Omega(\Psi)=0$, 
$I(\Psi)=0$], the event-horizon regularity condition can be regarded as 
an analog of a mixed-type von~Neumann--Dirichlet boundary condition 
linking the function $\Psi_0(\theta)$ to the first radial derivative 
$\partial\Psi/\partial r$ at the horizon (see \S~\ref {sec-zero-rotation}). 
Having noted this similarity between the horizon regularity condition and 
Dirichlet (or mixed-type) boundary conditions, I would like to stress 
that, at the same time, the {\it regularity} condition~(\ref{eq-EH-I-2}) 
cannot be regarded as an independently-imposed {\it boundary} condition.


\subsection{Regularity Condition at the Light Cylinder}
\label{subsec-LC}

In addition to the event horizon $\alpha=0$, there is another 
singular surface of equation~(\ref{eq-GS-2}), namely, the light 
cylinder $r=r_{\rm LC}(\theta)$ defined by the condition
\begin{equation}
\alpha^2 = R^2 \Omega^2(\Psi) \, .
\label{eq-LC-def}
\end{equation}

This is the surface where the electric field becomes
equal to the poloidal magnetic field and hence where
the toroidal velocity of the magnetic field lines,
$v_F=R \Omega/\alpha$, becomes equal to the speed of
light. This surface separates two regions: the outer
region, where $\alpha^2 > R^2\Omega^2$, and the inner
region, where $\alpha^2 < R^2\Omega^2$. The Grad--Shafranov
equation remains elliptic in both of these regions, but on 
the light-cylinder surface itself the coefficient in front 
of both the radial and the $\theta$ second derivatives becomes 
zero; hence this surface is a (regular) singular surface of the 
Grad--Shafranov equation.

Notice that the Grad--Shafranov equation admits solutions with
second-order derivatives that diverge at the light cylinder; when 
multiplied by $(\alpha^2-R^2\Omega^2)$, these derivatives could give 
a finite contribution that would be balanced by the other terms in the 
equation. Such solutions, however, would be characterized by a non-zero
jump of the first-order derivatives of $\Psi$ across the light cylinder; 
this would mean that magnetic field would not be continuous across the 
light cylinder, i.e., that a current sheet would develop along this 
surface. In the present study, I am interested in those magnetic field 
configurations that continue smoothly through the light cylinder without 
any such current sheets. This is because, as one can argue, any current 
sheets would dissipate due to resistive or other nonideal effects and 
cease to exist. Therefore, one can impose a light-cylinder regularity 
condition that states that $\Psi(r,\theta)$ should be a continuously 
differentiable function and hence the terms containing the second-order 
derivatives of $\Psi$ should give zero contribution at 
$r=r_{\rm LC}(\theta)$. Applying this regularity condition, together 
with~(\ref{eq-LC-def}), to equation~(\ref{eq-GS-2}), one then obtains 
the following equation:
\begin{equation}
II'(\Psi) = \Omega\Omega'(\Psi) R^2 |\nabla\Psi|^2 +
2\alpha^2 {\cot\theta\over{r^2}} {{\partial\Psi}\over{\partial\theta}} -
{\alpha^2\over r} (1-3\alpha^2) 
{{\partial\Psi}\over{\partial r}}, \qquad {\rm at}\ r=r_{\rm LC}(\theta)\, .
\label{eq-LC-regularity}
\end{equation}

Thus, the role of the light cylinder is that its presence enables one, 
upon imposing a regularity condition, to determine the remaining unknown 
function of~$\Psi$, namely, the poloidal current function~$I(\Psi)$. 
Equation~(\ref{eq-LC-regularity}) is just an explicit manifestation 
of this idea. In other words, of all the possible poloidal-current 
(and hence toroidal field) distributions~$I(\Psi)$, only the one 
that satisfies equation (\ref{eq-LC-regularity}) should give rise 
to a magnetic configuration without unphysical discontinuities.

As I have mentioned earlier, the field lines going from the surface 
of a rotating black hole to spatial infinity (such as those involved 
in the Blandford--Znajek process), should cross two light cylinders.
Correspondingly, there would be two regularity conditions. One then
should be able to use these two conditions to fix both unknown 
functions $I(\Psi)$ and $\Omega(\Psi)$; thus the problem would 
become fully determined without the need to invoke, as it is 
often done, a remote astrophysical load with uncertain physical 
properties. Whether this approach is physically valid, depends, 
however, on the physical properties of the putative particle-creation 
(or particle-injection) region. Such a region must necessarily be present 
somewhere between the two light-cylinder surfaces, because the particle
flux must be directed inward at the inner surface and outward at the 
outer surface. Even if the density of matter in this region is low
enough for the force-free approximation to be valid, the ideal-MHD
assumption may break down here. As a result, there could be a finite
voltage drop across this region (similar to the situation in the spark 
gap in pulsar magnetosphere) and, hence, $\Omega(\Psi)$ could have
different values on the two sides of this region. Then, the physics 
of the particle-creation region would have to provide the additional 
information necessary to fix the magnetospheric parameters (Beskin \& 
Kuznetsova 2000). It is conceivable, however, that the voltage drop, 
and hence the resulting jump in $\Omega$, would be small. Then, the
additional complications associated with the particle-creation region
could be ignored and the idea of using the two light-cylinder regularity 
conditions to determine the two functions $I(\Psi)$ and $\Omega(\Psi)$
would work. At the same time, the conditions of regularity could still 
be imposed at the event horizon and at infinity (but not the boundary 
conditions!). The regularity condition at infinity can be set if the 
domain under consideration extends to cylindrical distances much larger 
than the outer light-cylinder radius. If the amount of magnetic flux 
in the system is finite, this condition is conceptually similar to 
the regularity condition at the event horizon; it corresponds to an 
outgoing force-free electromagnetic wave and has the physical meaning 
of the force balance between the poloidal electric and the toroidal 
magnetic fields.


\section{Zero-rotation limit of the Grad--Shafranov equation}
\label{sec-zero-rotation}

Let us consider, as an important special case,
the zero-rotation limit of the Grad--Shafranov 
equation:
\begin{equation}
\Omega(\Psi) \equiv 0\, .
\label{eq-Omega=0}
\end{equation}

Even though this case is just a special limit of a more general
situation, there are some important differences between  this
case and the case with $\Omega\neq 0$. In particular, in the
$\Omega=0$ case the light-cylinder condition~(\ref{eq-LC-def}) 
becomes $\alpha=0$ and thus the light cylinder coincides with 
the event horizon. Therefore, as we shall see, the conditions 
set at the horizon can be used both to determine the poloidal-current 
function $I(\Psi)$ and as a regularity condition in the Grad--Shafranov 
equation. Indeed, upon substituting $\Omega(\Psi)=0$ into equation~(\ref
{eq-GS-ML}), one gets:
\begin{equation}
\alpha^2 \Delta^*\Psi \equiv
\alpha^2 \biggl[ \partial_r (\alpha^2 \partial_r \Psi) +
{{\sin\theta}\over r^2} \partial_\theta
\biggl({1\over{\sin\theta}}\partial_\theta \Psi\biggr)\biggr]=-II'(\Psi)\, .
\label{eq-GS-zero-rotation}
\end{equation}

At the event horizon $\alpha^2=0$, so, if one restricts oneself to
solutions that behave near $r=r_s$ according to~(\ref{eq-EH-Psi}),
the left-hand side (LHS) of this equation, and hence its RHS, vanish. 
Thus one concludes that in the absence of rotation the poloidal current 
is zero, $I(\Psi)\equiv 0$, and therefore the field is purely poloidal. 
Next, according to equation~(\ref{eq-E}), the electric field is 
identically zero and thus ${\bf j\times B}=0$. This means that
 the toroidal current must also be equal to zero and thus the 
no-rotation condition~(\ref{eq-Omega=0}) automatically leads 
to the absence of all electric currents in the magnetosphere. 
The magnetic field is then a vacuum field; upon setting the 
RHS of equation~(\ref{eq-GS-zero-rotation}) to zero and dividing 
this equation by $\alpha^2$, one obtains the following very simple 
linear equation that is valid everywhere outside the event horizon:
\begin{equation}
\Delta^*\Psi = \partial_r (\alpha^2 \partial_r \Psi) +
{{\sin\theta}\over r^2} \, \partial_\theta \biggl(
{1\over{\sin\theta}}\partial_\theta \Psi \biggr) =0\, 
\qquad r>r_s \, .
\label{eq-GS-vacuum}
\end{equation}
[General separable solutions of this equation, and the rich multipolar 
structure they form, have been derived and analyzed by Ghosh (2000).]

Now let us look at the behavior of the solution near the event horizon. 
Both indicial exponents of the linear equation~(\ref{eq-GS-vacuum}) are
equal to zero and hence one could expect some logarithmic terms to appear
in the asymptotic expansion near $r=r_s$.%
\footnote
{Several analytical solutions possessing such logarithmic singularities 
have been discussed, for the linear case $\Omega(\Psi)=0$, $I(\Psi)=I_0=
{\rm const}$, by Ghosh (2000).} 
However, the physical reasoning leading to regularity conditions~(\ref
{eq-EH-1})--(\ref{eq-EH-3}) is still valid in the zero-rotation case. 
Therefore, the asymptotic expansion should be regular: 
$\Psi(r,\theta)=\Psi_0(\theta)+\Psi_1(\theta)(r-r_s)+ 
\Psi_2(\theta)(r-r_s)^2+...$, as $r\rightarrow r_s$.
Upon substituting this expansion into equation~(\ref{eq-GS-vacuum}) 
and upon examining the lowest (zeroth) order terms in $(r-r_s)$, 
one derives the following event-horizon regularity condition:
\begin{equation}
r_s \Psi_1(\theta) + \Delta_\theta^*\Psi_0(\theta) = 0 \, ,
\label{eq-EH-vacuum-1}
\end{equation}
where I made use of $[\alpha^2(r)]'|_{r=r_s}=1/r_s$. 

Notice that equation~(\ref{eq-EH-vacuum-1}) can also be written in 
the form 
\begin{equation}
r_s {{\partial\Psi}\over{\partial r}} |_{r=r_s} +
\Delta_\theta^*\Psi|_{r=r_s} = 0 \, ,
\label{eq-EH-vacuum-2}
\end{equation}
which can also be obtained immediately from the Grad--Shafranov 
equation~(\ref{eq-GS-vacuum}) in the limit $r\rightarrow r_s$ 
simply by dropping the $\alpha^2\partial_r^2\Psi$ term.
Comparing equations~(\ref{eq-EH-II'}) and~(\ref{eq-EH-vacuum-2}) 
that represent the~Grad--Shafranov equation applied to the event 
horizon in the cases with and without rotation, respectively, one
can see that there is a remarkable difference between these two 
cases. Namely, in the case with rotation [$\Omega(\Psi)\neq 0$ 
and hence $I(\Psi)\neq 0$], the magnetic flux function on the 
event horizon, $\Psi_0(\theta)$, is determined from an ODE. 
That is, its only link to the magnetic field outside the horizon 
is provided by the two functions $\Omega(\Psi)$, $I(\Psi)$, but 
not by the radial derivatives of $\Psi$. In contrast, in the 
zero-rotation case, $\Psi(r,\theta)$ satisfies a PDE at the 
horizon; the connection between $\Psi_0(\theta)$ and the 
outside field $\Psi(r>r_s,\theta)$ cannot be maintained 
by $\Omega(\Psi)$ and $I(\Psi)$ because both of these two 
functions are identically zero in this case; instead, the 
connection is established by the radial derivative of~$\Psi$, 
as can be seen from equation~(\ref{eq-EH-vacuum-2}).

I have solved equation~(\ref{eq-GS-vacuum}) numerically, using 
the boundary conditions~(\ref{eq-bc-axis})--(\ref{eq-bc-gap}). 
More specifically, I have taken the inner radius of the disk 
$r_{\rm in}$ to be equal to the ISCO radius: $r_{\rm in}=r_{\rm ISCO}=6M$. 
In addition, I have restricted my consideration to only a single 
specific case,
\begin{equation}
\Psi_d(r)= {{r_{\rm ISCO}}\over r} = {6M\over r} \, ,
\label{eq-Psi_d}
\end{equation}
where I have set $\Psi_{\rm max}=\Psi_d(r_{\rm in})=1$.

When performing these calculations, I have used a relaxation procedure 
to solve the elliptic equation on a grid that was uniform in $\theta$ 
(60 gridpoints) and in the coordinate $x\equiv \sqrt{r_s/r}$ (100 
gridpoints). The relaxation procedure is a modification of the 
one used by Uzdensky~et~al. (2002) to study the non-relativistic 
force-free magnetospheres of magnetically-linked star--disk systems. 

Figure \ref{fig-contour-vacuum} shows the contour plot of the poloidal 
magnetic flux function for the zero-rotation case. Interestingly, the 
numerically-obtained function $\Psi_0^{(0)}(\theta)$ (the poloidal flux 
distribution on the horizon in the rotation-free case) is matched 
perfectly by a simple analytical expression describing the monopole
field:
\begin{equation}
\Psi_0^{(0)}(\theta)= 1- \cos{\theta} = 2\, \sin^2{\theta\over 2} \, .
\label{eq-Psi0-vacuum}
\end{equation}
This result is, in fact, not that surprising, considering that the 
equator boundary condition~(\ref{eq-bc-gap}) enforces a monopole-like 
field configuration all the way from the event horizon to the inner 
edge of the disk (see also Beskin \& Kuznetsova 2000).

Solution~(\ref{eq-Psi0-vacuum}) corresponds to a uniform radial 
magnetic field on the horizon:
\begin{equation}
B_r^{(0)}(r=r_s,\theta) = {1\over{r_s^2}}\, {1\over{\sin\theta}}\, 
{{d\Psi_0^{(0)}}\over{d\theta}} = {1\over{r_s^2}} = {1\over{4M^2}} \, .
\label{eq-Br-EH-vacuum}
\end{equation}
This implies that the assumption of the magnetic field being uniform 
at the horizon, adopted by Wang et al. (2002, 2003), is actually pretty 
reasonable, at least in the case of Schwarzschild black hole surrounded 
by a nonrotating (and, as we shall see in the next section, even by a 
Keplerian!) disk.

In addition, from equation~(\ref{eq-EH-vacuum-2}) one finds that 
in this case
\begin{equation}
{\partial\Psi^{(0)}\over{\partial r}} |_{r=r_s} \equiv 0 \, .
\end{equation}


\section {Slow-Rotation Case: Keplerian Disk}
\label{sec-slow}

Consider now a situation where the disk does rotate around
the black hole, but with a relatively small angular velocity,
in the sense that $R\Omega(\Psi) \ll 1$ everywhere. Then the 
light cylinder, whose position is determined by the condition~(\ref
{eq-LC-def}), has to lie very close to the event horizon:
\begin{equation}
\alpha_{\rm LC} = R\Omega(\Psi)|_{\rm LC} \ll 1 \qquad \Rightarrow 
\qquad r_{\rm LC} = {r_s\over{1-\alpha_{\rm LC}^2}} \simeq
r_s (1+\alpha_{\rm LC}^2) +O(\alpha_{\rm LC}^4) \, .
\label{eq-r_LC-slow}
\end{equation}

Next, since $\partial_r \Psi$ is finite at the event horizon 
according to~(\ref{eq-EH-Btheta=0}), one can estimate 
\begin{equation}
\Psi_{\rm LC}(\theta) \equiv \Psi[r=r_{LC}(\theta),\theta] \simeq 
\Psi_0(\theta)+(r_{\rm LC}-r_s)\,  
{{\partial\Psi}\over{\partial r}}|_{r=r_s} 
= \Psi_0(\theta) + O(\alpha_{\rm LC}^2) \, .
\label{eq-Psi_LC-slow}
\end{equation}

In addition, one can approximate $R_{\rm LC}(\theta) \equiv
r_{\rm LC}(\theta) \sin\theta = r_s \sin\theta + O(\alpha_{\rm LC}^2)$
on the RHS of equation~(\ref{eq-LC-def}) and thus obtain:
\begin{equation}
\alpha_{\rm LC}(\theta) = r_s \sin\theta \, \Omega[\Psi_0(\theta)] +
O(\alpha_{\rm LC}^3) \, .
\label{eq-alpha_LC-slow}
\end{equation}
Thus, in the slow-rotation case, the location of the light cylinder 
with respect to the event horizon becomes immediately determined in 
terms of $\Psi_0(\theta)$ and $\Omega(\Psi)$.

At this point I would like to digress to argue that, in fact, 
a Keplerian disk%
\footnote
{The assumption that the disk rotates with the Keplerian angular 
velocity is justified when one can neglect the action of magnetic 
forces on the disk on the rotation-period time scale, i.e., when 
$V_{A,d} \ll V_K$.} 
can be regarded as slowly-rotating to a very good degree of 
approximation. Indeed, a Keplerian disk is characterized by
\begin{equation}
\Omega_K (r) \equiv {{M^{1/2}}\over{r^{3/2}}}\, , \qquad 
r \geq r_{\rm ISCO} \equiv 6M \, ,
\label{eq-Omega-Keplerian}
\end{equation}
The maximum value $\Omega_{\rm max}$ of $\Omega_K(r)$ is achieved 
at $r_{\rm ISCO}=6M$: $\Omega_{\rm max} = \Omega_K(r_{\rm ISCO}) =
6^{-3/2} M^{-1} \simeq 0.068/M$. In this paper I am considering 
magnetic configuration in which the field line passing through 
the inner edge of the disk lies entirely in the equatorial plane 
$\theta=\pi/2$ and spans the entire plunging region $(r_s\leq r 
\leq r_{\rm ISCO}; \theta=\pi/2)$. This line then intersects the 
light cylinder at $\alpha_{\rm LC}(\theta=\pi/2)\simeq r_s\Omega_{\rm max}=
1/3\sqrt{6}\simeq 0.136\ll 1$; correspondingly, using equation~(\ref
{eq-r_LC-slow}), $r_{\rm LC}(\pi/2)-r_s\simeq M/27 \ll r_s$. For all 
other field lines [with $r_0(\Psi)>r_{\rm ISCO}$], the corresponding 
values of $\alpha_{\rm LC}$ and of the difference $r_{\rm LC}-r_s$ 
are even less than these, because of the smaller values of both 
$\Omega(\Psi)$ and $\sin\theta$. One thus sees that the most 
physically-interesting case of a Keplerian disk can indeed 
be described very well by the slow-rotation limit.

The next important simplification in the slow-rotation limit comes from 
the fact that, according to equation~(\ref{eq-EH-I-2}), the poloidal-current 
function $I(\Psi)$ is proportional to $\Omega(\Psi)$ and is thus also small. 
This enables one to regard both $\Omega(\Psi)$ and $I(\Psi)$ as giving rise 
to only small perturbations in the Grad--Shafranov equation~(\ref{eq-GS-2}). 
Accordingly, one can expect the solution $\Psi(r,\theta)$ of this equation 
to be approximated very closely by the solution $\Psi^{(0)}(r,\theta)$ 
of equation~(\ref{eq-GS-vacuum}) describing the zero-rotation case. 
In particular, this means that the event-horizon flux distribution 
$\Psi_0(\theta)$ is very close to $\Psi_0^{(0)}(\theta)=1-\cos\theta$.

To verify the validity of these claims, I have solved the full 
Grad--Shafranov equation~(\ref{eq-GS-2}) (for a Keplerian disk) 
numerically, subject to the same boundary conditions (\ref
{eq-bc-axis})--(\ref{eq-bc-gap}), (\ref{eq-Psi_d}). In this 
numerical solution I have had to locate the light cylinder 
and have used the light cylinder regularity condition~(\ref
{eq-LC-regularity}) to determine the function $I(\Psi)$ iteratively, 
until convergence was achieved. The numerical procedure that I have 
used has, once again, been a modification of the procedure used and 
described by Uzdensky~et~al. (2002).

The contour plot of the poloidal magnetic flux function $\Psi(r,\theta)$
for the case of Keplerian disk is presented in Figure~\ref{fig-contour}.
One can easily see that the difference between this case and the 
zero-rotation case (presented in Figure~\ref{fig-contour-vacuum}) 
is indeed almost imperceptible. One also finds that the numerically-obtained
function $\Psi_0(\theta)$ coincides perfectly with the analytical
expression~(\ref{eq-Psi0-vacuum}). Thus one concludes that, when 
studying the magnetosphere of a Schwarzschild black hole magnetically
linked to a slowly-rotating (e.g., Keplerian) disk, one can indeed 
replace the exact solution $\Psi_0(\theta)$ of the full nonlinear 
force-free Grad--Shafranov equation~(\ref{eq-GS-2}) by the more 
readily computable function $\Psi_0^{(0)}(\theta)\equiv\Psi_0(\theta)
|_{\Omega=0}$, which comes from the solution of the simpler linear 
PDE~(\ref{eq-GS-vacuum}). This conclusion is very important as one 
now no longer needs to use the event-horizon condition~(\ref{eq-EH-I-2}) 
for the determination of $\Psi_0(\theta)$, as required by our usual, 
proper procedure. Consequently, this condition is now freed up and 
one can use it for a direct determination of the function~$I(\Psi)$, 
instead of having to use for this purpose the complicated light-cylinder 
regularity condition~(\ref{eq-LC-regularity}). As an example, I shall now 
demonstrate this streamlined procedure for the case of Keplerian disk. 

With the disk magnetic flux distribution $\Psi_d(r)$ given by 
equation~(\ref{eq-Psi_d}) and with the Keplerian rotation velocity 
$\Omega_K(r)$ given by~(\ref{eq-Omega-Keplerian}), the function
$\Omega(\Psi)$ can be expressed explicitly as
\begin{equation}
\Omega(\Psi) = {1\over M}\, \biggl({\Psi\over 6}\biggr)^{3/2}\, ,
\label{eq-Omega-of-Psi}
\end{equation}

Together, the two functions $\Psi_d(r)$ and $\Psi_0(\theta)$ provide 
a mapping relation between the disk surface and the event horizon.
Thus, using expression~(\ref{eq-Psi0-vacuum}) for the horizon
distribution $\Psi_0(\theta)$, one immediately obtains:
\begin{equation}
\Omega[\Psi_0(\theta)] = {1\over{3^{3/2} M}}\, \sin^3{\theta\over 2} \, ,
\end{equation}
\begin{equation}
\alpha_{\rm LC}(\theta) = {2\over{3^{3/2}}}\, \sin\theta \, 
\sin^3{\theta\over 2} \, ,
\label{eq-alpha_LC-Keplerian}
\end{equation}
and hence, using equation~(\ref{eq-EH-I-2}), one gets
\begin{equation}
I[\Psi_0(\theta)] = {1\over{3^{3/2}M}}\, \sin^2\theta \, 
\sin^3{\theta\over 2} \, ,
\end{equation}
Combining this formula with the expression~(\ref{eq-Psi0-vacuum}) 
for $\Psi_0(\theta)$, one can finally express $I$ as a function of~$\Psi$:
\begin{equation}
I(\Psi) = {{\Psi^{5/2}(2-\Psi)}\over{6^{3/2}M}} \, .
\label{eq-I-of-Psi}
\end{equation}

All of these expressions agree perfectly with my numerical results for 
the Keplerian disk.


\section{Conclusions and Discussion of Future Plans}
\label{sec-conclusions}

In this paper I have studied an axisymmetric stationary
force-free magnetosphere of a Schwarzschild black hole
in the presence of a thin ideally-conducting accretion 
disk. Such a magnetosphere is described by the Grad--Shafranov
equation --- a second-order elliptic nonlinear Partial 
Differential Equation for the poloidal magnetic flux
function~$\Psi$. The problem is further complicated 
by the presence in this equation of two functions 
of~$\Psi$, the angular velocity of the magnetic field 
lines $\Omega(\Psi)$ and the poloidal current $I(\Psi)$, 
that need to be somehow specified for the problem to be 
fully determined. 

I have restricted my consideration to the so-called
Magnetically-Coupled configuration in which all the 
magnetic field lines that emerge from the hole's 
(stretched) event horizon connect to the disk surface. 
In this case, the Grad--Shafranov equation possesses 
two regular singular surfaces, the event horizon and 
the inner light cylinder. Correspondingly, I have set 
two regularity conditions, one at each surface. I have 
used the event-horizon regularity condition to determine
the horizon's magnetic flux distribution $\Psi_0(\theta)$ and the 
light-cylinder regularity condition to fix the function~$I(\Psi)$. 
In addition, I have prescribed two functions at the disk surface: 
the poloidal flux distribution $\Psi_d(r)$, which I have used as 
a boundary condition for $\Psi$ at the equatorial plane, and the 
disk angular velocity $\Omega_d(r)$. Under the assumption that 
the disk is infinitely conducting, the magnetic field lines in 
a steady state have to rotate with the angular velocity of their 
disk footpoints; thus, the functions $\Omega_d(r)$ and $\Psi_d(r)$ 
together determine $\Omega(\Psi)$, i.e., the second function of 
$\Psi$ present in the Grad--Shafranov equation. With all these 
conditions specified, and with $\Psi$ set equal to zero along 
the rotation axis $\theta=0$ and at infinity, the problem has 
now been fully determined mathematically.

I have then obtained numerical solutions of the problem
for two important specific cases. The first one is the 
case of a nonrotating disk, $\Omega(\Psi)=0=I(\Psi)$.
The Grad--Shafranov equation is greatly simplified and
becomes linear in this case. By solving it numerically,
I have found that the radial magnetic field is uniform
on the black hole's event horizon, corresponding to the
split-monopole horizon flux distribution $\Psi_0(\theta)=
1-\cos\theta$.

The second case I have considered is the case of a Keplerian disk.
I first have argued that this case can be analyzed in the slow-rotation 
limit of the Grad--Shafranov equation, $R\Omega\ll c$. In this limit, 
the inner light cylinder lies very close to the horizon, i.e., 
$\alpha_{LC} \ll 1$. In addition, the poloidal current $I(\Psi)$ 
is also small and hence the poloidal-field structure of the magnetosphere, 
described by $\Psi(r,\theta)$, is in fact very close to that corresponding 
to the zero-rotation case. In particular, this means that one can use the 
zero-rotation result $\Psi_0(\theta)=1-\cos\theta$ to obtain exact 
analytical expressions for the functions describing the slow-rotation, 
e.g., Keplerian, case, such as the location $\alpha_{LC}(\theta)$ of 
the light cylinder and the function $I(\Psi)$. In addition to 
deriving these expressions, I have solved the full nonlinear 
problem for the Keplerian disk numerically, without making 
the slow-rotation approximation. I have found my analytical 
predictions to be in perfect agreement with the numerical 
results.

As I have discussed in the Introduction, the present work,
dealing with a Schwarzschild black hole, should be viewed 
simply as a first step in a larger project. More relevant 
and more physically-interesting is, of course, the case of
the magnetosphere of a Kerr black hole. In this case, the 
magnetic connection can lead to the transfer of energy and 
angular momentum from the rapidly-rotating black hole to 
the disk, thereby changing the disk's observable spectra 
(Li 2000, 2001, 2002, 2003). In addition, one may expect 
that the toroidal magnetic field, generated due to the 
twisting of the poloidal magnetic field lines by the rapidly 
spinning black hole, will exert a strong outward pressure on 
the poloidal field; this, in turn, may lead to a significant 
inflation and even a partial opening of the magnetic field. 
Such a process, if it does occur, would be very similar to 
the analogous process of field-line inflation and opening 
due to toroidal-field pressure known to take place in 
differentially-rotating force-free magnetospheres of 
magnetically-linked star--disk systems (e.g., van~Ballegooijen 1994; 
Lovelace~et~al. 1995; Uzdensky~et~al. 2002; Uzdensky 2002a,b).
In the case of an accreting Kerr black hole, this process would 
be extremely important, as it would lead to a simultaneous, hybrid 
action of the Magnetic-Coupling process (on the closed field lines) 
and the Blandford-Znajek process (on the open field lines). Solving 
the Grad--Shafranov equation should then give us the location of the 
separatrix between the open and closed field-line regions and hence 
an estimate of the relative importance of these two processes as a 
function of the black-hole spin parameter~$a$.

These arguments provide the motivation for extending
the present work to the Kerr case in the near future. 
In addition to purely technical complications, simply 
due to a larger number of terms in the equations, an 
analysis of the Kerr case will probably also require 
the development of a proper treatment for the open field 
lines. This includes, for example, the combined use of the 
inner and outer light-cylinder regularity conditions to fix 
the two functions $\Omega(\Psi)$ and $I(\Psi)$ and also 
prescribing the appropriate conditions at infinity (see 
the discussion at the end of \S~\ref{sec-regularity}).

Another direction for future research has to do with a more
realistic description of the disk. Indeed, in the present 
paper I assumed that the disk is perfectly conducting and 
arbitrarily prescribed the magnetic flux distribution, $\Psi_d(r)$, 
on its surface [in particular, I took $\Psi_d(r)\sim 1/r$]. Whereas 
a thin disk, even when it is turbulent, can indeed be considered a 
perfect conductor {\it on the rotation-period time scale}, in the 
longer term this is not so. If the disk is turbulent (due to the 
magneto-rotational instability, for example), it will have some 
effective turbulent magnetic diffusivity. If the large-scale poloidal 
field approaches such a disk at a finite angle [i.e., if $(B_r/B_z)_d 
=O(1)$], this effective diffusivity will lead to a relatively fast 
resistive slippage of the magnetic footpoints in the radial direction 
(with the velocity of the order of $v_{\rm turb}\gg v_{\rm accretion}$), 
and thus to a relatively rapid rearrangement of the flux distribution 
$\Psi_d(r)$. A quasi-steady state (on time scales much longer than the 
rotation period) can be established only if the large-scale poloidal
magnetic field is nearly perpendicular to the surface of the (turbulent) 
disk. Thus, I believe that the von~Neumann disk boundary condition 
$\partial_\theta\Psi(r>r_{\rm in},\theta=\pi/2)=0$ is physically 
better motivated than the Dirichlet boundary condition $\Psi(r>r_{\rm in},
\theta=\pi/2)=\Psi_d(r)$ adopted in the present paper.

I would like to thank Vasilii Beskin, Arieh K{\"o}nigl, B.~C.~Low, 
Leonid Malyshkin, Vladimir Pariev, and Brian Punsly for their 
encouragement and fruitful discussions. This research was supported 
by the National Science Foundation under Grant No.~PHY99-07949.


\section{References}

Begelman, M.~C., Blandford, R.~D., \& Rees, M.~J. 1984, 
Rev. Mod. Phys., 56, 255

Bender, C.~M., \& Orszag, S.~A. 1978, Advanced Mathematical Methods 
for Scientists and Engineers, McGrow-Hill, Inc., New York

Beskin, V.~S., \& Par'ev, V.~I. 1993, Phys. Uspekhi, 36, 529

Beskin, V.~S. 1997, Phys. Uspekhi, 40, 659

Beskin, V.~S., \& Kuznetsova, I.~V. 2000, Nuovo Cimento, 115, 795;
preprint (astro-ph/0004021)

Blandford, R.~D. 1999, in Astrophysical Disks:
An EC Summer School, ed. J.~A.~Sellwood \& J.~Goodman 
(San Francisco: ASP), ASP Conf. Ser. 160, 265; preprint 
(astro-ph/9902001)

Blandford, R.~D. 2000, Phil. Trans. R. Soc. Lond. A, 358, 811;
preprint (astro-ph/0001499)

Blandford, R.~D., \& Znajek, R.~L. 1977, MNRAS, 179, 433 (BZ77)

Contopoulos, I., Kazanas, D., \& Fendt, C. 1999, ApJ, 511, 351

Damour, T. 1978, Phys. Rev. D, 18, 3589

Fendt, C. 1997, A\&A, 319, 1025

Gammie, C.~F. 1999, ApJ, 522, L57

Ghosh, P. 2000, MNRAS, 315, 89

Gruzinov, A. 1999, preprint (astro-ph/9908101)

Hirotani, K., Takahashi, M., Nitta, S.-Y., \& Tomimatsu, A. 1992,
ApJ, 386, 455

Komissarov, S.~S. 2001, MNRAS, 326, L41

Krolik, J.~H. 1999, Active Galactic Nuclei:
From The Central Black Hole To The Galactic Environment
(Princeton: Princeton Univ. Press)

Li, L.-X. 2000, ApJ, 533, L115

Li, L.-X. 2001, in X-ray Emission from Accretion onto 
Black Holes, ed. T.~Yaqoob \& J.~H.~Krolik, JHU/LHEA Workshop, 
June 20-23, 2001

Li, L.-X. 2002, A\&A, 392, 469

Li, L.-X. 2003, Phys. Rev. D, 67, 044007

Lovelace, R.~V.~E., Romanova, M.~M., \& Bisnovatyi-Kogan, G.~S.
1995, MNRAS, 275, 244 

Macdonald, D., \& Thorne, K.~S. 1982, MNRAS, 198, 345 (MT82) 

Macdonald, D.~A. 1984, MNRAS, 211, 313

Mobarry, C.~M., \& Lovelace, R.~V.~E. 1986, ApJ, 309, 455

Nitta, S.-Y., Takahashi, M., \& Tomimatsu, A. 1991, Phys. Rev. D, 44, 2295

Okamoto, I. 1974, MNRAS, 167, 457

Phinney, E.~S. 1983, in Astrophysical Jets, ed. A.~Ferrari \&
A.~G.Pacholczyk (Dordrecht: Reidel), 201

Punlsy, B. 1989, Phys. Rev. D, 40, 3834

Punsly, B. 2001, Black Hole Gravitohydromagnetics (Berlin: Springer)

Punlsy, B., \& Coroniti, F.~V. 1990, ApJ, 350, 518

Thorne, K.~S., Price, R.~H., \& Macdonald, D.~A. 1986,
Black Holes: The Membrane Paradigm (New Haven: Yale Univ. Press)

Uzdensky, D.~A., K{\"o}nigl, A., \& Litwin, C. 2002, ApJ, 565, 1191

Uzdensky, D.~A., 2002a, ApJ, 572, 432

Uzdensky, D.~A., 2002b, ApJ, 574, 1011

Uzdensky, D.~A., 2003, ApJ, accepted; preprint (astro-ph/0305288)

van~Ballegooijen, A.~A. 1994, Space Sci. Rev., 68, 299

Wang, D.~X., Xiao, K., \& Lei, W.~H. 2002, MNRAS, 335, 655

Wang, D.-X., Lei, W.~H., \& Ma, R.-Y. 2003, MNRAS, 342, 851

Znajek, R.~L. 1977, MNRAS, 179, 457

Znajek, R.~L. 1978, MNRAS, 185, 833


\clearpage

\begin{figure}
\plotone{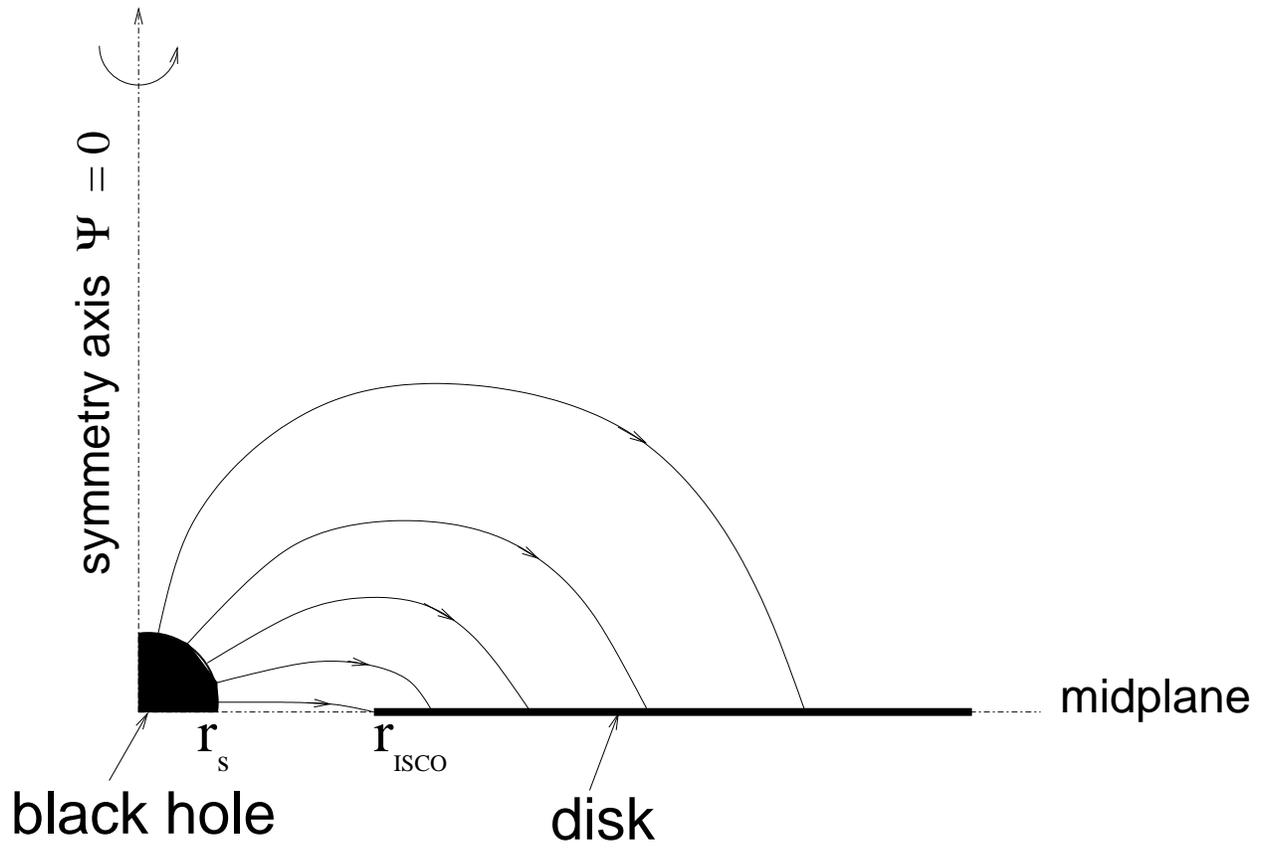}
\figcaption{Schematic drawing of an axisymmetric linked 
Black Hole -- Disk magnetosphere.
\label{fig-geometry}}
\end{figure}

\clearpage

\begin{figure}
\plotone{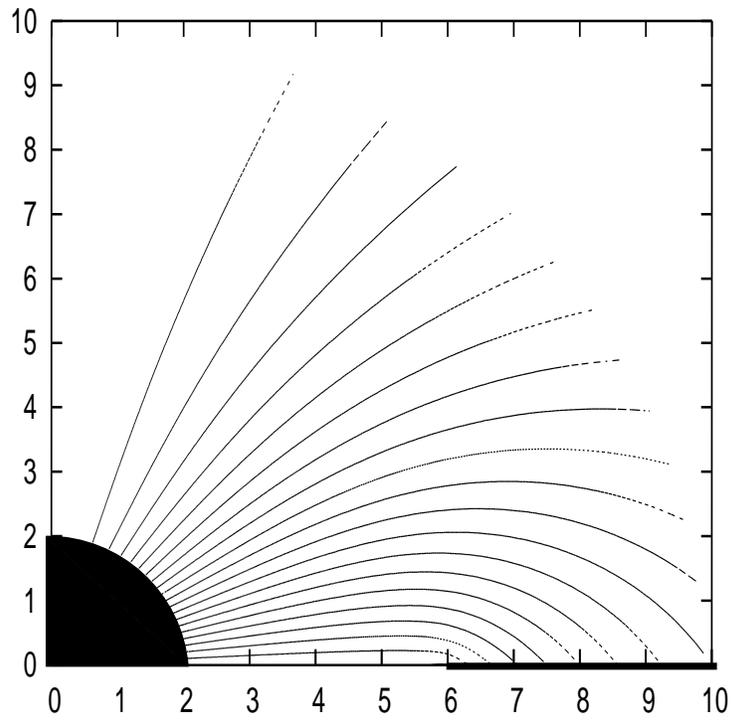}
\figcaption{Contour plot of the magnetic flux function $\Psi(r,\theta)$
in the zero-rotation case, $\Omega(\Psi)=0=I(\Psi)$.
\label{fig-contour-vacuum}}
\end{figure}

\clearpage

\begin{figure}
\plotone{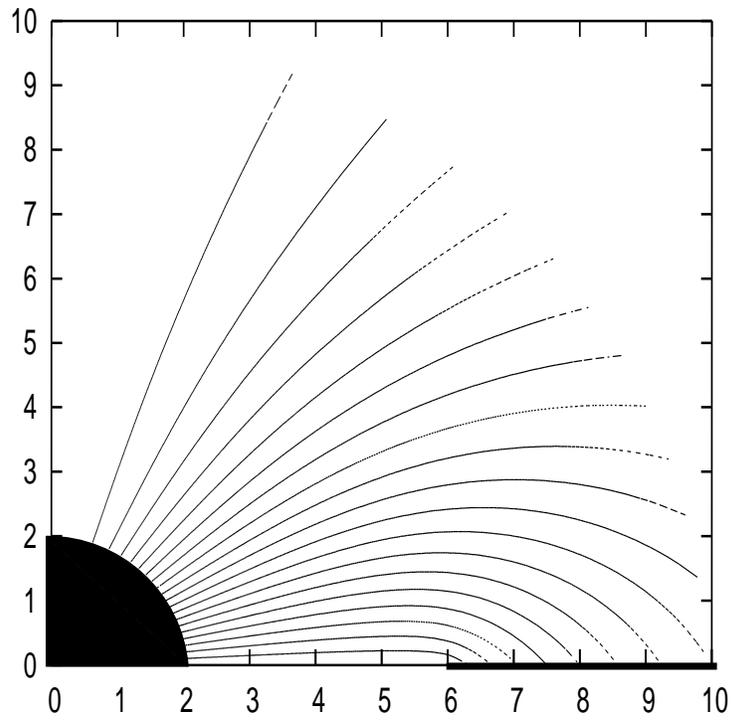}
\figcaption{Contour plot of the magnetic flux function $\Psi(r,\theta)$
for the case of Keplerian disk. 
\label{fig-contour}}
\end{figure}

\end{document}